\documentclass[aps,prd,twocolumn,showpacs]{revtex4}

\usepackage{amsfonts,amssymb,amsmath,latexsym,graphicx}

\newcommand{\pd}{\partial}

\def\Kc{{\cal K}}
\def\Pc{{\cal P}}

\begin{document}

\title{Energy Flow from Open to Closed Strings in a Toy Model of Rolling Tachyon}

\author{Liudmila Joukovskaya}
\email{ljoukov@mi.ras.ru}
\affiliation{%
Steklov Mathematical Institute\\
Russian Academy of Sciences\\
Gubkin St. 8, 119991 Moscow, Russia}

\author{Yaroslav Volovich}
\email{Yaroslav.Volovich@msi.vxu.se}
\affiliation{%
Moscow State University, 119899 Moscow, Russia\\
MSI, V\"axj\"o University, SE-351 95, Sweden}

\begin{abstract}
We study the toy model of interacting open and closed string tachyons which
demonstrates some interesting properties of the unstable D-brane decay scenario.
We compute a stress tensor of the system and study the energy and pressure
dynamics. We show that the total energy of the system is conserved.
We separate the stress tensor into two parts corresponding to open and closed
strings and study the energy flow from open to closed strings.
The two vacua of the system could be interpreted as corresponding to
the unstable D-brane background and the open string tachyon vacuum.
We study how for spatially homogenous solution interpolating between
these two vacua the total energy of the open string dissipates to the closed string.
\end{abstract}


\maketitle

\newpage
\section{Introduction}

Physical process of the unstable D-brane decay has recently attracted a lot of
attention
\cite{0203211,0207107,Kluson,0209122,Yang,VolYa,AJ,0303139,0304163,0304213,%
0305011,0305159,0305191,VSVYaV,Oh}.
This process has been investigated in
frameworks of boundary CFT \cite{0203211,0303139} and within open string field
theory (OSFT) \cite{0207107,Kluson,AJ,0304213}. To work within  OSFT a
level-truncated scheme has been  usually employed. Dynamical equations in this
approach are rather interesting from mathematical point of view since they
contain  an infinite number of derivatives. These equations could be
represented in the integral form that corresponds to a nonlocal interaction
found in string theories. For detailed analysis of such types of equations see
for example \cite{BFOW,0207107,Yang,VolYa,AJ}, for some rigorous mathematical
results in this field see \cite{VSVYaV}. Despite of these mathematical
difficulties several interesting results where obtained. In particular special
time dependent solutions of classical equations of motion where constructed in
p-adic string theory \cite{VVZ,Kh,Fr-rev} in papers \cite{BFOW,0207107} and in
SSFT in papers \cite{VolYa,AJ}. These solutions describe the tachyon field
which starts from the unstable vacuum of the system with non-zero
velocity and tends to the stable one reaching it at infinite
time. The D-brane's tension at the unstable vacuum is expected to be at
its maximum and vanishing while moving to the stable one.

An interesting point in the current development of the
unstable D-brane decay is the disagreement between the conformal
field theory considerations and the results obtained in the
level-truncated open cubic string field theory
approximations.
This gave rise to an investigation of open-closed interacting string
models where it was proposed that one has to take into account
the fact that the D-brane's energy dissipates to the closed string
while reaching the resulting stable vacuum.

In this paper we consider a model with two interacting tachyon fields
recently proposed in \cite{Oh}.
The model could be seen as a simplification
of level-truncated Open-Closed String Field Theory (OCSFT) where
several terms are omitted even on the first nontrivial level.
Although this model could be considered only as a simplified model of
OCSFT it has a time dependent solution interpolating between
two vacua of the system.
This solution starts from the first vacuum interpreted
as an unstable D-brane background and tends to the second one where
the D-brane disappears \cite{Oh}.
Here we show that for this solution the conserved energy dissipates
from the open string tachyon to the closed one.
This coincides with the expected behavior of the D-brane's energy.

The paper is organized as follows.
In Section \ref{sec:model} we describe the model and study its dependance
on the parameters.
Then in Section \ref{sec:str-t} we compute a stress tensor
of the model and separate the terms corresponding to open and
closed tachyons.
Finally in Section \ref{sec:en-pr} we study energy and pressure dynamics.
First, we provide a direct prove of the energy convergence.
Then using the separation of terms in the stress tensor we study
energy flow from open to closed strings.
We show how the energy of open string dissipates to closed one.
At last we show a picture of the pressure dynamics.
The pressure is negative and vanishes at infinite time.

\section{The Toy Model for Open and Closed String Tachyons}
\label{sec:model}

We study a toy model of open-closed string tachyons recently
proposed in \cite{Oh}. In the units with $\alpha^\prime=1$ the
corresponding action takes the form
\begin{multline}
\label{act}
S=\int d^Dx
\left[
\frac{1}{2}\phi \Box \phi+\frac{1}{2}\phi^2+
\frac{1}{2}\psi \Box \psi+2\psi^2\right.\\
\left.
-\frac{1}{3}\tilde{\phi}^3+c_2
\tilde{\phi}\tilde{\psi}-\tilde{\phi}^2\tilde{\psi}
\right],
\end{multline}
where $\phi$ and $\psi$ are interpreted as open and closed string tachyon
fields respectively \cite{Oh}, and one defines
\begin{subequations}
\begin{equation}
\label{int1}
\tilde{\phi}(x)\equiv e^{k \Box}\phi(x),
\end{equation}
\begin{equation}
\label{int2}
\tilde{\psi}(x)\equiv e^{m \Box}\psi(x),
\end{equation}
\end{subequations}
$k=m=\log 2$ with the differential operator $e^{a\Box}$ is understood
as a series expansion
\begin{equation}
\label{serie}
e^{a\Box}=\sum_{n=0}^\infty \frac{a^n \Box^n}{n!},
\end{equation}
(about an appearance of nonlocal terms in SFT see reviews \cite{0102085,0111208,0301094})
and $c_2$ is a constant which is discussed below. The space-time is flat with
the metric $\eta^{\mu\nu}=\mbox{diag}(-1,1,\ldots,1)$, so that the D'Alamber operator
takes the form $\Box=-\frac{\pd^2}{\pd t^2}+\nabla^2$.
The equations of motion corresponding to the action (\ref{act}) are the following
\begin{subequations}
\begin{equation}
\label{eom1}
\Box \phi +\phi-e^{k\Box}\tilde{\phi}^2+c_2e^{k\Box}\tilde{\psi}-
2e^{k\Box}(\tilde{\phi}\tilde{\psi})=0
\end{equation}
\begin{equation}
\label{eom2}
\Box \psi +4\psi+c_2e^{m\Box}\tilde{\phi}-
e^{m\Box}\tilde{\phi}^2=0
\end{equation}
\end{subequations}

Let us consider spatially homogeneous configurations. In this case the
equations of motion (\ref{eom1})-(\ref{eom2}) take the following form
\begin{subequations}
\begin{equation}
\label{eom1t=0}
(-\pd^2+1)e^{2k\pd^2}\tilde{\phi}-\tilde{\phi}^2+c_2\tilde{\psi}-2\tilde{\phi}\tilde{\psi}=0
\end{equation}
\begin{equation}
\label{eom2t=0}
(-\pd^2+4)e^{2m\pd^2}\tilde{\psi}+c_2\tilde{\phi}-\tilde{\phi}^2=0,
\end{equation}
\end{subequations}
where the time derivative is denoted by $\partial=\frac{d}{dt}$ and
$\tilde{\phi}=\tilde{\phi}(t)$, $\tilde{\psi}=\tilde{\psi}(t)$.

Equations of motion (\ref{eom1t=0})-(\ref{eom2t=0}) contain
a pseudo-dif\-feren\-tial operator of the form $e^{a\pd^2}$, it is defined as
\begin{equation}
\label{K-def}
e^{a\pd^2}\varphi(t)=\int \Kc_a(t-\tau) \varphi(\tau) d\tau,
\end{equation}
where $\Kc_a$ is a Gaussian kernel given by
\begin{equation}
\label{ker}
\Kc_a(x)=\frac{1}{\sqrt{4\pi a}} e^{-\frac{x^2}{4a}}
\end{equation}
The integral operator (\ref{K-def}) is well defined for $a > 0$, see
\cite{BFOW,0207107,VolYa,VSVYaV} for detailed analysis of equations with such operators.

The equations of motion have tree vacua solutions which could be easily found
by the following procedure. First we solves
the equation (\ref{eom2}) for constant $\phi$, $\psi$
with respect to $\psi$. This gives $\psi=(\phi^2-c_2\phi)/4$.
Now substituting it to (\ref{eom1}) and using the fact that the fields are
constant we get an equation $\phi(2\phi^2+(4-3c_2)\phi+c_2^2-4)=0$
which has three solutions
$$
\phi_0=0,~~\phi_{1,2}=\frac{1}{4}(3c_2-4\pm\sqrt{c_2^2-24c_2+48})
$$
these are the vacua of the system in terms of $c_2$.

The potential of the system has the form
\begin{equation}
\label{V-pot}
V=\frac{1}{8}\phi^4+
  \left(-\frac{1}{4}c_2+\frac{1}{3}\right)\phi^3+
  \left(\frac{1}{8}c_2^2-\frac{1}{2}\right)\phi^2
\end{equation}

Now we want to fix a constant $c_2$ in such a way that there are two distinct
vacua with the same energy, i.e. we solve equations $V(\phi_i)=V(\phi_j)$,
$i,j=0,1,2$ with respect to $c_2$ and check that for this value of $c_2$
$\phi_i\neq\phi_j$.
The first step gives us the following values of $c_2$
$$
c_2=\pm 2,~~c_2=\frac{13}{6},~~c_2=\frac{4}{3},~~c_2=4(3\pm\sqrt{6})
$$
Although only for the values $c_2=\frac{13}{6}$ and $c_2=\frac{4}{3}$
there are two {\it distinct} vacua with the same energy.

\begin{figure}
\centering
\includegraphics[width=4.2cm]{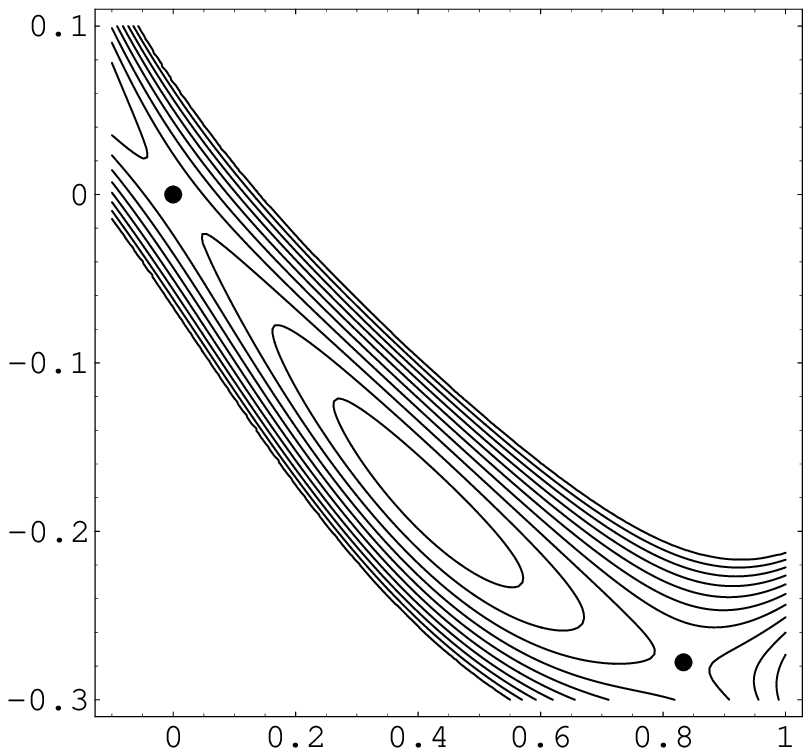}
\includegraphics[width=4.2cm]{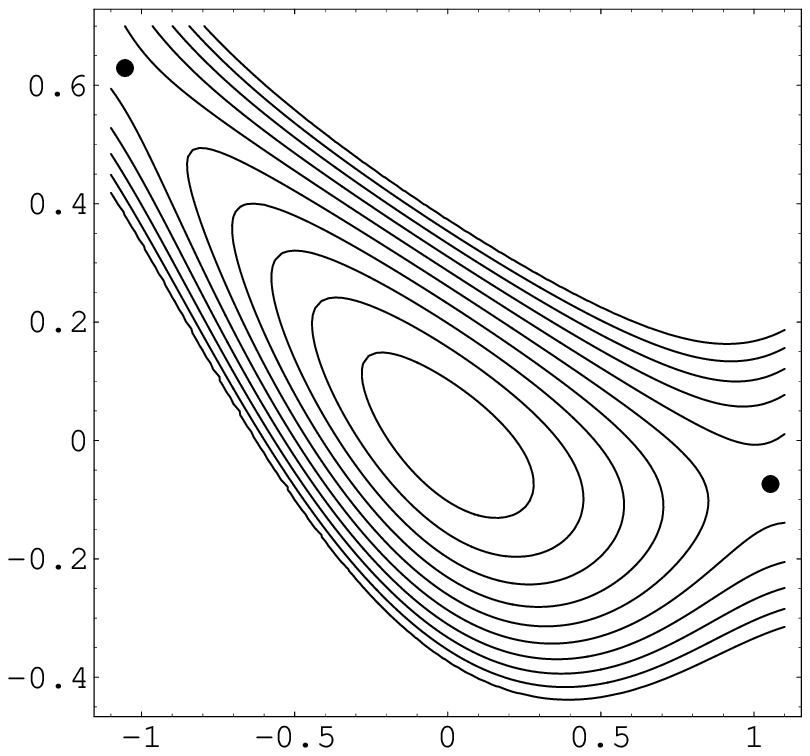}
\caption{
Equipotential lines of the effective mechanical potential (\ref{V-eff}).
For the case $c_2=\frac{13}{6}$ (left) there are two vacua with the same energy:
$(\tilde{\phi}=0,\tilde{\psi}=0)$ and $(\tilde{\phi}=\frac{5}{6},\tilde{\psi}=-\frac{5}{18})$,
for the case $c_2=\frac{4}{3}$ (right) there are again two distinct vacua
with the same energy:
$(\tilde{\phi}=\pm\frac{\sqrt{10}}{3},\tilde{\psi}=\frac{5\mp 2\sqrt{10}}{18})$.
On the figures the corresponding vacua are marked with thick dots.}
\label{fig:pot}
\end{figure}

\begin{figure}
\centering
\includegraphics[width=4.2cm]{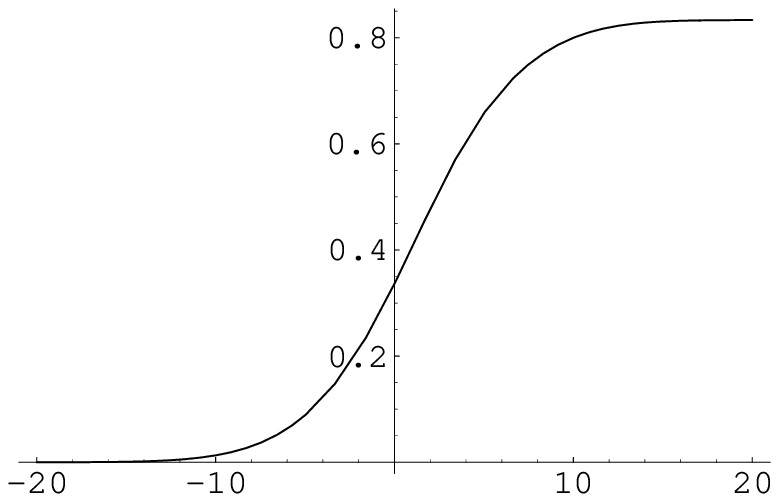}
\includegraphics[width=4.2cm]{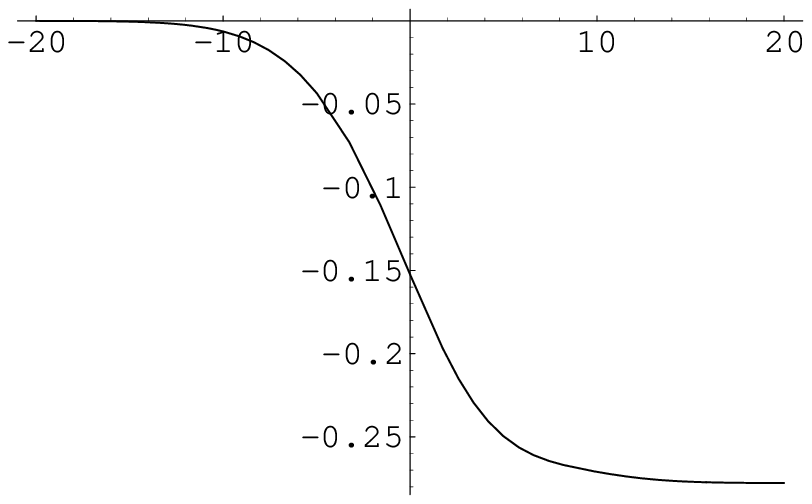}
\caption{Spatially homogeneous solutions for the case $c_2=\frac{13}{6}$ constructed
using the iterative procedure (\ref{iter-proc}): $\phi(t)$ (left) and $\psi(t)$ (right).}
\label{fig:sols}
\end{figure}

In \cite{Oh} a space homogeneous configuration was constructed numerically for
the case $c_2=\frac{13}{6}$ using the following iterative procedure
\begin{equation}
\label{iter-proc}
\begin{split}
\tilde{\phi}_{n+1}&=\frac{1}{c_2}(\tilde{\phi}_n^2-\Pc_4\tilde{\psi}_n),\\
\tilde{\psi}_{n+1}&=\frac{1}{c_2}
  (-\Pc_1\tilde{\phi}_n+\tilde{\phi}_n^2+2\tilde{\phi}_n\tilde{\psi}_n),
\end{split}
\end{equation}
where $n=1,2,\ldots$ and the initial configuration is taken as
$\phi_0(t)=\frac{5}{6}\theta(t)$, $\psi_0(t)=-\frac{5}{18}\theta(t)$,
where $\theta(t)$ is a step function being $1$ for $t>0$ and $0$ otherwise.
The operator $\Pc_r=(-\pd^2+r)e^{2k\pd^2}$ could be presented in
the fully integral form by differentiating the kernel (\ref{ker})
\begin{equation}
\label{Pr_op}
(\Pc_r\varphi)(t)=\int (-\frac{d^2}{dt^2}+r) K_{2k}(t-\tau)\varphi(\tau)d\tau
\end{equation}
The results of the iterative procedure (\ref{iter-proc}) are presented on
Fig.\ref{fig:sols}.

In order to support this numerical results let us study the equations
(\ref{eom1t=0})-(\ref{eom2t=0}) neglecting high order derivatives.
This reduces the original equations to the following system
(see (\ref{int1})-(\ref{int2}))
\begin{subequations}
\begin{equation}
\label{ec1}
(2k-1)\pd^2\tilde{\phi}+\tilde{\phi}-\tilde{\phi}^2+c_2\tilde{\psi}-2\tilde{\phi}\tilde{\psi}=0
\end{equation}
\begin{equation}
\label{ec2}
(8m-1)\pd^2\tilde{\psi}+4\tilde{\psi}+c_2\tilde{\phi}-\tilde{\phi}^2=0
\end{equation}
\end{subequations}
Equations (\ref{ec1})-(\ref{ec2}) could be seen as describing a mechanical system
with the following potential
\begin{equation}
\label{V-eff}
V_{\text{eff}}(\tilde{\phi},\tilde{\psi})=
  \frac{1}{2}\tilde{\phi}^2+2\tilde{\psi}^2-
  \frac{1}{3}\tilde{\phi}^3+c_2\tilde{\phi}\tilde{\psi}-\tilde{\phi}^2\tilde{\psi}.
\end{equation}
We will call potential obtained in such a way an effective mechanical potential.
Note that
\begin{equation}
\label{pot-rev}
V_{\text{eff}}\big(\tilde{\phi},\frac{1}{4}(\tilde{\phi}^2-c_2\tilde{\phi})\big)=-V(\tilde{\phi})
\end{equation}
this reflects the fact that interacting terms contain second order derivatives and thus
contribute to the kinetic term. We have this effect of potential-flipping only if
$k>\frac12$ and $m>\frac18$ (we have this in our model since $k=m=2$),
i.e. the effective ``masses'' are positive.

The described construction of an effective mechanical potential provide good intuition
about the existence of rolling-type solutions. Indeed the shape of
$V_{\text{eff}}$ (Fig.\ref{fig:pot}) points
(at least for the case $c_2=\frac{13}{6}$) to the existence of kink-type solutions for
the reduced system (\ref{ec1})-(\ref{ec2}) and this usually a good sign for the
original nonlocal system to have a rolling-type solution \cite{AJ,VolYa}.
The direct proof of existence of rolling-type solutions for our system might be
rather nontrivial. Indeed the properties of the integral kernel (\ref{Pr_op}) does
not allow us to use the same approach as was used in \cite{VSVYaV}.

Although the iterative procedure (\ref{iter-proc}) demonstrates a good numerical
convergence for $c_2=\frac{13}{6}$ it rapidly diverges for $c_2=\frac{4}{3}$ where
the initial functions where taken to be ``stepping'' between the corresponding
vacua
\begin{equation}
\label{it043}
\phi_0=\pm\frac{\sqrt{10}}{3},~~
\psi_0=\frac{5\mp 2\sqrt{10}}{18},
\end{equation}
It is left unclear whether kink-type solution exists even for the reduced system
(\ref{ec1})-(\ref{ec2}) for this value of $c_2$. Our numerical investigations show that
the arch at the bottom of the potential for this case (see Fig.\ref{fig:pot}, part right)
prevents the trajectory started from one endpoint (\ref{it043}) to reach the second.

Below we will refer to the spatially homogeneous solution for the case
$c_2=\frac{13}{6}$ displayed on Fig.\ref{fig:sols} \cite{Oh}.

\section{Stress Tensor}
\label{sec:str-t}

Let us compute a stress tensor for the model and
present it in the form with the separated terms corresponding to
open and closed string.

We use the definition of a stress tensor from general relativity
\begin{equation}
\label{T-def}
T_{\alpha\beta}=\frac{2}{\sqrt{-g}}\frac{\delta S}{\delta g_{\alpha\beta}}
\end{equation}
We make the action (\ref{act}) invariant under space-time transformations
by including metric in the action
\begin{multline}
\label{action}
S=\int d^D x \sqrt{-g}
\left[
\frac{1}{2}\phi\Box\phi+\frac{1}{2}\phi^2+
\frac{1}{2}\psi\Box\psi+2\psi^2\right.\\
\left.
-\frac{1}{3}\tilde{\phi}^3+c_2\tilde{\phi}\tilde{\psi}-\tilde{\phi}^2\tilde{\psi}
\right]
\end{multline}
where the D'Alamber operator $\Box$ is covariant
$$
\Box=\frac{1}{\sqrt{-g}} \partial_{\mu}\sqrt{-g}g^{\mu \nu}\partial_{\nu}
$$

While performing a variation of the action (\ref{action}) according to
(\ref{T-def}) we will have to deal with the following expressions
\begin{equation}
\label{constr1}
\int d^D y ~f(y)\frac{\delta \Box_{y}}{\delta g_{\alpha \beta}(x)}h(y)
\end{equation}
and
\begin{equation}
\label{constr2}
\int d^D y ~f(y)\frac{\delta \tilde{\chi}(y)}{\delta g_{\alpha \beta}(x)},
~~~\tilde{\chi}(y)\equiv e^{n \Box}\chi(y),
\end{equation}
here and below the index $y$ denotes derivation with respect to $y$.

Let us now consider these expressions in more detail \cite{Yang,AJ}.
Using the following equality
$$
\frac{\delta\sqrt{-g(y)}}{\delta g^{\alpha \beta}(x)}=-\frac{1}{2}\sqrt{-g(y)}
g_{\alpha \beta}(y)\delta(x-y)
$$
we compute
\begin{multline}
\label{varsq}
\frac{\delta \square_y}{\delta g^{\alpha\beta}(x)}
=\frac{g_{\alpha\beta}}{2}\delta(x-y)\square_y
\\
+\frac{1}{\sqrt{-g(y)}}\partial_{y_{\mu}}
\Big[\frac{-\sqrt{-g(y)}}{2}\delta(x-y)g_{\alpha\beta}(y)g^{\mu \nu}(y)
\\
+\sqrt{-g(y)}\delta^{\mu \nu}_{\alpha\beta}\delta(x-y)\Big]\partial_{y_{\nu}}
\end{multline}
Now for the expression (\ref{constr1}) we obtain
\begin{multline}
\label{comp1}
\int d^D y ~f(y)\frac{\delta \Box_{y}}{\delta g_{\alpha \beta}(x)}h(y)
=\int d^D y ~f(y)
\\
\times\Big[
    \frac{g_{\alpha\beta}}{2}\delta(x-y)\square_y
    +\frac{1}{\sqrt{-g}}\partial_{y_{\mu}}
    \Big(
        \frac{-\sqrt{-g}}{2}\delta(x-y)
\\
        \times g_{\alpha\beta}g^{\mu\nu}
        +\sqrt{-g}\delta^{\mu \nu}_{\alpha\beta} \delta(x-y)
    \Big)
    \partial_{ y_{\nu}}
\Big]
h(y)
\end{multline}

Now let us consider the expression (\ref{constr2}). First let us rewrite it
in the following form
$$
\int d^D y ~f(y)\frac{\delta \tilde{\chi}(y)}{\delta g_{\alpha \beta}(x)}=
\int d^D y ~f(y)\frac{\delta e^{n \Box }}{\delta g_{\alpha \beta}(x)} \chi(y)
$$
Using (\ref{varsq}) and the equality \cite{Yang}
$$
\frac{\delta \hat{A}}{\delta g^{\alpha \beta} (x)}=
\int_{0}^{1} d \rho e^{\rho \hat{A}}( \frac{ \delta \hat{A}}{\delta g^{\alpha \beta} (x)})
e^{(1-\rho) \hat{A}},
$$
we obtain
\begin{multline}
\label{comp2}
\int d^D y ~f(y)\frac{\delta \tilde{\chi}(y)}{\delta g_{\alpha \beta}(x)}
\\
=\frac{g_{\alpha \beta}}{2}
\left[
    n\int_0^1 d\rho(e^{n\rho \square}f )(\square e^{n(1-\rho)\square}\chi)+
\right.
\\
\left.
    +n\int_0^1d\rho(\partial_{\mu}e^{n\rho \square}f)
    (\partial^{\mu} e^{n(1-\rho)\square}\chi)
\right]
\\
-n\int_0^1 d\rho(
\partial_{\alpha}e^{n\rho \square}f )(\partial_{\beta} e^{n(1-\rho)\square} \chi)
\end{multline}

Now using (\ref{comp1}) and (\ref{comp2}) we are able to compute the stress tensor
of the system
\begin{equation}
\label{str-t}
T_{\alpha\beta}=T_{\alpha\beta}^{\phi}+T_{\alpha\beta}^{\psi},
\end{equation}
where the part corresponding to the open string tachyon is
\begin{multline*}
T_{\alpha\beta}^{\phi}=-g_{\alpha \beta}
\left[ -\frac{1}{2}\partial_\mu\phi\partial^\mu\phi+
\frac{1}{2}\phi^2-\frac{1}{3}\tilde{\phi}^3\right]
-\partial_\alpha\phi\partial_\beta\phi
\\
-g_{\alpha\beta}
\left[k\int_0^1 d\rho (e^{-(2-\rho)k\Box}(\Box+1)\tilde{\phi})
(\Box e^{-k\rho \Box}\tilde{\phi})\right.
\\
\left.+ k\int_0^1 d\rho
(\partial_{\mu}e^{-(2-\rho)k
\Box}(\Box+1)\tilde{\phi})(\partial^{\mu}
e^{-k\rho \Box}\tilde{\phi})\right]
\\
+2k\int_0^1 d\rho (\partial_{\alpha}e^{-(2-\rho)k\Box}
(\Box+1)\tilde{\phi})(\partial_{\beta} e^{-k\rho \Box}\tilde{\phi}),
\end{multline*}
and the part corresponding to the closed string tachyon is given by
\begin{multline*}
T_{\alpha\beta}^{\psi}=
-g_{\alpha\beta}
\left[-\frac{1}{2}\partial_\mu\psi\partial^\mu\psi
+2\psi^2-\tilde{\psi}e^{-2m\Box}(\Box+4)\tilde{\psi}\right]
\\
-\partial_\alpha\psi\partial_\beta\psi
\\
-g_{\alpha\beta} \left[m \int_0^1 d\rho (e^{-(2-\rho)m
\Box}(\Box+4)\tilde{\psi})(\Box e^{-m\rho \Box}\tilde{\psi})\right.
\\
\left.
  +m\int_0^1 d\rho
(\partial_{\mu}e^{-(2-\rho)m
\Box}(\Box+4)\tilde{\psi})(\partial^{\mu}
e^{-m\rho \Box}\tilde{\psi})\right]
\\
+2 m\int_0^1 d\rho (\partial_{\alpha} e^{-(2-\rho)m
\Box}(\Box+4)\tilde{\psi})(\partial_{\beta} e^{-m\rho \Box}\tilde{\psi}),
\end{multline*}
where we used the equations of motion.
We see that the terms of the stress tensor corresponding to open and closed
strings are separated. This allows us to study the energy flow from open to closed
strings.

\section{Energy and Pressure Dynamics}
\label{sec:en-pr}

The energy of the system is defined in terms of the stress tensor (\ref{str-t})
\begin{equation}
\label{E-def}
E=T^0_0
\end{equation}
As it was shown in the previous section it is possible to separate the terms
of the stress tensor corresponding to open and closed strings.
Thus the energy of the system (\ref{E-def}) could also be written as a sum
of energies of open and closed strings.

For spatially homogeneous configurations the energy takes the form
\begin{equation}
E(t)=E_{\phi}(t)+E_{\psi}(t),
\end{equation}
where the term corresponding to open string is given by
\begin{multline*}
E_{\phi}(t)=-\frac{1}{2}\phi^2+\frac{1}{2}(\partial \phi)^2+\frac{1}{3}\tilde{\phi}^3\\
+k \int_0^1 d\rho (e^{(2-\rho)k
\partial^2}\{-\partial^2+1\}\tilde{\phi})\overleftrightarrow{\partial}(e^{k\rho
\partial^2}\partial\tilde{\phi}),
\end{multline*}
and the term corresponding to closed string is
\begin{multline*}
E_{\psi}(t)=-2\psi^2+\frac{1}{2}(\partial\psi)^2+
\tilde{\psi}\left[(-\partial^2 +4)e^{2m \partial^2}\tilde{\psi}\right]\\
+m\int_0^1 d\rho (e^{(2-\rho)m
\partial^2}\{-\partial^2+4\}\tilde{\psi})\overleftrightarrow{\partial}(
e^{m\rho\partial^2}\partial\tilde{\psi)},
\end{multline*}
here and below $A\overleftrightarrow{\partial}B=A\partial B-B\partial A$.
\begin{figure}
\centering
\includegraphics[width=5cm]{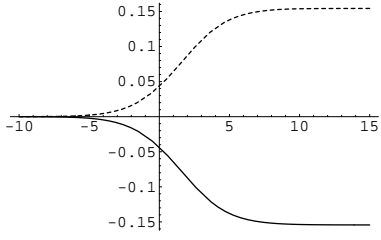}
\caption{The the energy of closed string $E_\psi$ (dashed line)
and the energy of open string $E_\phi$ (solid line).}
\label{fig:en}
\end{figure}

One can immediately see that this energy conserves
\begin{multline*}
\frac{dE(t)}{dt}=-\phi \partial \phi-4\psi \partial \psi+\tilde{\phi}^2\partial\tilde{\phi}+
\partial\phi\partial^2\phi+\partial\psi\partial^2\psi
\\
+\partial\tilde{\psi}\left[(-\partial^2+4)e^{2m\partial^2}\tilde{\psi}\right]+
\tilde{\psi}\left[(-\partial^2+4)e^{2m\partial^2}\partial\tilde{\psi}\right]
\\
+k \int_0^1 d\rho (e^{(2-\rho)k
\partial^2}\{-\partial^2+1\}\tilde{\phi})\overleftrightarrow{\partial^2}(e^{k\rho
\partial^2}\partial\tilde{\phi})
\\
+m \int_0^1 d\rho (e^{(2-\rho)m
\partial^2}\{-\partial^2+4\}\tilde{\psi})\overleftrightarrow{\partial^2}
(e^{m\rho\partial^2}\partial\tilde{\psi)},
\end{multline*}
using the identity \cite{AJ}
\begin{equation}
m\int_{0}^{1} d \rho (e^{m\rho\partial^2}\varphi)\overleftrightarrow{\partial^2}
(e^{m(1-\rho)\partial^2}\phi)=
\varphi\overleftrightarrow{e^{m\partial^2}}\phi,
\end{equation}
we get
\begin{multline*}
\frac{dE(t)}{dt}
=-\phi \partial \phi-4\psi \partial \psi+\tilde{\phi}^2\partial\tilde{\phi}+
\partial\phi\partial^2\phi+\partial\psi\partial^2\psi
\\
+\partial\tilde{\psi}
\left[(-\partial^2+4)e^{2m\partial^2}\tilde{\psi}\right]+
\tilde{\psi}
\left[(-\partial^2+4)e^{2m\partial^2}\partial\tilde{\psi}\right]
\\
-(\partial\tilde{\phi})\overleftrightarrow{e^{k\partial^2}}(\{-\partial^2+1\}\phi)
-(\partial\tilde{\psi})\overleftrightarrow{e^{m\partial^2}}(\{-\partial^2+4\}\psi)
\end{multline*}
or simplifying
$$
=\tilde{\phi}^2\partial\tilde{\phi}
-\partial\tilde{\phi}\left[\{-\partial^2+1\}e^{k\partial^2}\phi\right]
+\tilde{\psi}\left[\{-\partial^2+4\}e^{m\partial^2}\pd\psi\right]
$$
now using the equation of motion (\ref{eom2}) we get
\begin{multline*}
=\tilde{\phi}^2 \partial\tilde{\phi}
-\partial\tilde{\phi}\left[\{-\partial^2+1\}e^{k\partial^2}\phi\right]
+\tilde{\psi}\partial(\tilde{\phi}^2-c_2 \tilde{\phi})
\\
=\partial\tilde{\phi}\left[-(-\partial^2+1)e^{k\partial^2}\phi+\tilde{\phi}^2-
c_2\tilde{\psi}+2\tilde{\phi}\tilde{\psi} \right]\equiv 0
\end{multline*}
We see that the time derivative of the energy is zero on the equations of motion
and thus the total energy is conserved. We also see that on infinities the
energy is zero and thus $E(t)=0$ or $E_{\phi}(t)=-E_{\psi}(t)$ at all times.
The energy flow from open to closed string for the solution
described in the section \ref{sec:model} is presented on Fig.\ref{fig:en}.
We see that the energy contained in the open string tachyon dissipates to
the closed string tachyon that could be interpreted as a transformation
of the unstable D-brane's energy to the energy of closed string \cite{Oh}.

Let us now investigate the pressure of the system \cite{0207107,Yang,AJ}.
The pressure is defined in terms of the stress tensor as
\begin{equation}
p_i=-T^i _i~~(\mbox{no summation in}~i).
\end{equation}

\begin{figure}
\centering
\includegraphics[width=5cm]{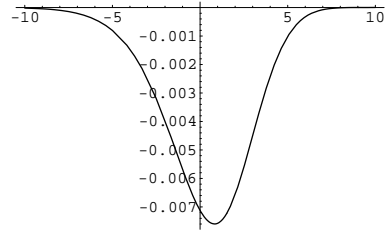}
\caption{The pressure of the system (\ref{press}) is negative and vanishes
on infinities.}
\label{fig:press}
\end{figure}

\begin{figure}
\centering
\includegraphics[width=5cm]{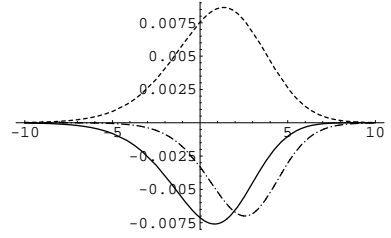}
\caption{The positive term $(\pd\phi)^2+(\pd\psi)^2$ of the pressure (dashed line),
the negative terms (dashed-dotted line) and the total pressure (solid line).
We see that there is a nontrivial compensation which leads to negative total pressure.}
\label{fig:parts}
\end{figure}

Since we consider spatially homogeneous configurations we will omit the vector index $i$.
The pressure dynamics is given by
\begin{multline}
\label{press}
p(t)=-E(t)+(\partial\phi)^2+(\partial\psi)^2\\
-2k \int_0^1 d\rho (e^{(2-\rho)k
\partial^2}\{-\partial^2+1\}\partial\tilde{\phi})
(e^{k\rho\partial^2}\partial\tilde{\phi})\\
-2 m \int_0^1 d\rho (e^{(2-\rho)m
\partial^2}\{-\partial^2+4\}\partial\tilde{\psi})
(e^{m\rho\partial^2}\partial\tilde{\psi)}
\end{multline}
The pressure dynamics is presented on Fig.\ref{fig:press}.
Since the total energy is zero and the fields are constant at infinities
we see from (\ref{press}) that the pressure vanishes as $|t|\to\infty$.
The pressure contains two positive terms with only second-order derivatives
of $\phi$ and $\psi$ and two negative terms with high-order derivatives
of $\tilde{\phi}$ and $\tilde{\psi}$. These terms are presented separately on
Fig.\ref{fig:parts}. We see that there is a nontrivial compensation of the
positive terms making the total pressure being negative.

Let us note that the form (\ref{press}) is rather nontrivial from computational point
of view. In particular
we where able to present the last two terms in such a way that the first
exponent is well defined in terms of the integral operator (\ref{K-def}) for all
$0\leqslant\rho\leqslant 1$ while there is no such possibility for the second
exponent in both terms --  the Gaussian kernel diverges at $\rho=0$. Although
for $\rho$ near to $0$ we where able to use the series expansion (\ref{serie})
without loss of precision.

\section{Conclusion}

In this paper a simplified model of the unstable D-brane decay
in the open-closed string field theory was studied. We have computed a stress
tensor of the system and presented it as a sum of two terms corresponding
to open and closed strings. We have obtained the energy of the system
and proved its convergence. The separation of energies of open and closed
strings allowed us to study the energy flow from open to closed strings.
We have considered a time-dependant solution interpolating between two
vacua of the system for the case $c_2=\frac{13}{6}$. We have showed
that the energy of open string dissipates to the closed string.
The question of existence of such interpolating solution for the case
$c_2=\frac{4}{3}$ is left open.
The next important step would be to generalize these results to the full
level-truncated OCSFT and study the D-brane's decay there.

\begin{acknowledgments}
The authors are grateful to
I.Ya.~Aref'eva for attention and discussions.
L.J. was financially supported by the grant RFBR-02-01-00695
and by Dynasty Foundation and International Center for Fundamental Physics in Moscow,
Ya.V. was financially supported by the grants
RFBR-02-01-01084 and RFBR-MAS-03-01-06466.
All numerical computations in this work where performed
using a C++ library developed by the authors, for visualization
we have used the {\it Mathematica} system.
\end{acknowledgments}

\end{document}